\documentclass[]{aa}  

\usepackage{graphicx}
\usepackage{txfonts}
\usepackage{hyperref}
\usepackage[utf8]{inputenc}
\usepackage[T1]{fontenc}
\usepackage{siunitx} 

\begin{document} 

    \title{Using photometric redshift data to improve the detection of
        galactic filaments with the Bisous model}
    \titlerunning{Using photometric redshift data with the Bisous model}

    \author{M. M. Muru\inst{1} \and E. Tempel\inst{1,2}}

    \institute{
    Tartu Observatory, University of Tartu, Observatooriumi 1, 61602 Tõravere, Estonia\\
    \email{moorits.mihkel.muru@ut.ee}
    \and
    Estonian Academy of Sciences, Kohtu 6, 10130 Tallinn, Estonia
    }

   \date{\today}
 
  \abstract
   {Filament finders are limited, among other things, by the abundance of spectroscopic redshift data.
   This limits the sky areas and depth where we can detect the filamentary network.}
   {As there are proportionally more photometric redshift data than spectroscopic, we aim to use data with photometric redshifts to improve and expand the areas where we can detect the large-scale structure of the Universe.
   The Bisous model is a filament finder that uses only the galaxy positions.
   We present a proof of concept, showing that the Bisous filament finder can improve the detected filamentary network with photometric redshift data.}
   {We created mock data from the \textsc{MultiDark-Galaxies} catalogue.
   Galaxies with spectroscopic redshifts were given exact positions from the simulation.
   Galaxies with photometric redshifts were given uncertainties along one coordinate.
   The errors were generated with different Gaussian distributions for different samples.
   We sample the photometric galaxy positions for each Bisous run based on the uncertainty distribution.
   In some runs, the sampled positions are closer to the true positions and produce persistent filaments; other runs produce noise, which is suppressed in the post-processing.}
   {There are three different types of samples: spectroscopic only, photometric only, and mixed samples of galaxies with photometric and spectroscopic redshifts.
   In photometric-only samples, the larger the uncertainty for photometric redshifts, the fewer filaments are detected, and the filaments strongly align along the line of sight.
   Using mixed samples improves the number of filaments detected and decreases the alignment bias of those filaments.
   The results are compared against the full spectroscopic sample.
   The recall for photometric-only samples depends heavily on the size of uncertainty and dropped close to 20\%; for mixed samples, the recall stayed between 40\% and 80\%.
   The false discovery rate stayed below 5\% in every sample tested in this work.
   Mixed samples showed better results than corresponding photometric-only or spectroscopic-only samples for every uncertainty size and number of spectroscopic galaxies in mixed samples.}
   {Mixed samples of galaxies with photometric and spectroscopic redshifts help us to improve and extend the large-scale structure further than possible with only spectroscopic samples.
   Although the uncertainty sizes tested in this work are smaller than those for the available photometric data, upcoming surveys, such as J-PAS, will achieve sufficiently small uncertainties to be useful for large-scale structure detection.}


   \maketitle

\section{Introduction} \label{sect:introduction}

The galaxy distribution in the observable Universe is not homogeneous but has a structure that is dictated by matter distribution and gravitational forces.
The large-scale structure of the Universe defines the environment galaxies reside in and has a wide range of effects on the properties of those galaxies; for example, the orientation of galaxies in relation to the filaments \citep{LeePen2000, AragonCalvo2007, TempelLibeskind2013, Ganeshaiah2019, Kraljic2020}, the satellite distribution around larger galaxies \citep{Knebe2004, Zenter2005, Tempel2015, Wang2020}, the elliptical-to-spiral ratio, and the star formation rate \citep{Alpaslan2015, Kuutma2017}.

Usually, the large-scale structure is divided into four types of substructures \citep{Libeskind18}.
The densest and most compact are galaxy clusters that host many gravitationally bound galaxies and are called knots in the large-scale structure context.
The clusters are connected by chains of galaxies called filaments that populate the intricate cosmic web.
Between clusters and filaments are large under-dense volumes named voids encapsulated in sheets of filaments called walls or sheets.

There are many different approaches to detecting the different large-scale structure elements.
Usually, the methods use either the relative positions of the galaxies themselves or different scalar and tensor fields derived from galaxy positions and properties from observational surveys or simulation data.
For example, the NEXUS+ model \citep{Cautun2013} uses the Hessian of the shear tensor field.
Models that use galaxy positions also employ different approaches.
For example, DisPerSE \citep{Sousbie2011} uses mass estimates and identifies the cosmic web using topological features of the mass distribution, and the Bisous model \citep{Tempel16} uses the distribution of the galaxies and marked point process with interactions.
\citet{Libeskind18} gives an overview and a brief comparison of 12 different methods to detect the large-scale structure elements.

The accuracy of these finders depends on the completeness and accuracy of the data.
The best results are obtained from the simulations where galaxy positions and properties are accurate and complete phase-space information is available.
When using data from surveys where the data are incomplete and have uncertainties, and phase-space information is derived from those observations, the resulting cosmic web maps deteriorate.
Some methods are better suited for observational data, but all methods are limited by the completeness and accuracy of spectroscopic redshift data.
The current largest spectroscopic survey is the Sloan Digital Sky Survey (SDSS, \citealt{SDSS_III, SDSS_DR12}), which covers \SI{7221}{deg\squared} of the sky.
There are upcoming large spectroscopic surveys such as the 4-metre Multi-Object Spectroscopic Telescope (4MOST, \citealt{4MOST2019}) surveys and the Dark Energy Spectroscopic Instrument (DESI, \citealt{DESI}) Bright Galaxy Survey (BGS, \citealt{DESI_BGS}).
Future surveys will cover larger areas but will still be limited by depth and completeness.

Data with photometric redshifts are much more abundant than the spectroscopic counterpart, as redshifts can be measured in bulk.
For example, SDSS Data Release 12 has 100 times more photometric redshifts than spectroscopic ones \citep{Beck2016}.
The upcoming J-PAS \citep{JPAS} will observe the sky in 54 narrowband and three broadband filters and is designed to measure the redshifts for a large number of galaxies with a precision of \(\sigma_z \lesssim 0.003(1+z)\).
This precision is comparable to low-resolution spectroscopic surveys and enables wider use of photometric redshift data for applications that require positions of galaxies, such as large-scale structure detection.

In this paper, we use the Bisous filaments finder, which is developed to detect filaments from observational data.
The Bisous model only needs the galaxy distribution and uses geometric methods and the marked point process with interactions to detect the cosmic web \citep{Tempel14, Tempel16}.
Bisous has been successfully used in many works, such as \citet{Nevalainen15}, \citet{Kuutma2017}, \citet{Ganeshaiah2019}, and \citet{Tuominen2021}.
\citet{Kruuse2019} show a significant positive correlation between the distribution of photometric galaxies and the Bisous filaments, which suggests that the Bisous model could be able to use photometric data to improve the detection of filaments.

In this study, we present a proof of concept that data with photometric redshifts can be used to improve the detection of the filamentary network.
For this, we take a simple approach to use data with significant uncertainties in position along one axis with the Bisous model.
We generate mock data with photometric and spectroscopic redshifts from a simulation and use samples with only photometric redshifts, mixed samples of photometric and spectroscopic redshifts, and, for comparison and benchmarking, also samples with only spectroscopic redshifts.
Using Bisous results from the full spectroscopic redshift data as a reference, we study the recall and false discovery rate of the Bisous runs on different samples.
Further aspects of interest are whether or not using data with photometric redshift produces biases in the filaments and the maximum size of uncertainties that Bisous can handle while still improving the filamentary network.

The structure of the paper is as follows.
In Sect. \ref{sect:data}, we describe the simulation we used to create the mock data and samples in this study.
In Sect. \ref{sect:bisous}, we describe the Bisous filament finder and our method to use data with photometric redshifts.
In Sect. \ref{sect:results}, we present the results from different samples.
A discussion of the results, problems, possible improvements, and future applications is presented in Sect. \ref{sect:discussion} and conclusions are outlined in Sect. \ref{sect:conclusions}.

\section{Data} \label{sect:data}

\subsection{Simulation data}

The analysis in this paper is based on simulated mock data.
For the mock data set, we used the galaxy catalogue \textsc{MultiDark-Galaxies} which is based on the \textsc{MultiDark-Planck 2} (MDPL2\footnote{\url{https://www.cosmosim.org/cms/simulations/mdpl2/}}, \citealt{Klypin16}) simulation with the \textsc{Sag} semi-analytic model for galaxies described in \cite{Knebe18}.
The MDPL2 simulation is based on a dark-matter-only flat \(\Lambda\) cold dark matter (\(\Lambda\)CDM) model with \textsc{Planck} cosmological parameters: \(\Omega_\mathrm{m} = 0.307, \ \Omega_\mathrm{B} = 0.048, \ \Omega_\Lambda = 0.693, \ \sigma_8 = 0.823, \ n_s = 0.96\), and \(h = 0.678\) \citep{Planck15}.
The box size is \SI{1000}{\mathit{h}^{-1}\,Mpc} (\SI{1475.6}{Mpc}) with \(3840^3\) particles with a mass resolution of \(m_\mathrm{p} = \SI{1.51e9}{\mathit{h}^{-1}\,\mathit{M_\odot}}\) per dark matter particle.

This work uses a smaller box of the whole simulation with a side of \SI{250}{Mpc} to have a sufficiently large sample size for statistical analysis but a sufficiently small volume to limit the calculation time of the Bisous filament finder (see Section~\ref{sect:bisous}) applied to the data.
We used a magnitude limit of -20.0 in the SDSS r-band to have galaxy number density similar to observations (for comparison, see \citealt{Muru21}).
This cut leaves us with \num{181411} galaxies in a box with a side of \SI{250}{Mpc}, and the galaxy number density is \SI{0.0116}{Mpc\tothe{-3}}.

\subsection{Photometric redshift mock data}

As the distance measures from spectroscopic surveys are relatively precise, the spectroscopic redshift mock data are simply data with exact positions from the simulation, but in order to generate photometric redshift mock data we have to introduce photometric redshift uncertainties to them.
The simulation data positions form a cube for which we take two axes to represent the sky plane, and the coordinates represent the sky coordinates and therefore have no extra uncertainty, and one axis represents the line of sight.
We added a random error to the line of sight coordinate of each galaxy.
For simplicity, all the coordinates are given in megaparsecs (\si{Mpc}), and the errors do not scale with distance.

The random errors for the line-of-sight axis are generated with a Gaussian distribution (\(\mathcal{N}(x, \sigma^2)\)) with different standard deviation (\(\sigma\)) values for different samples.
Within one sample, the standard deviation value is constant.
For this study, we used six different standard deviation values of \SIlist{1;2;3;5;7;10}{Mpc}.
We also created mixed samples of galaxies with spectroscopic and photometric distances in different proportions and with different photometric uncertainties.
This is to emulate a realistic situation where one would start with an observational catalogue of spectroscopic targets and include photometric targets to improve the detection of the large-scale structure.
Different mixed samples have \SIlist{10;20;30;40;50}{\%} of the brightest galaxies as spectroscopic galaxies; the rest are photometric galaxies with uncertainties generated with \(\sigma = \) \SIlist[list-pair-separator = { or }]{5;10}{Mpc}.
This means that a chosen percentage of the brightest galaxies have exact positions and other galaxies have photometric uncertainties in the line of sight axis, such as distance.

\begin{figure*}
    \centering
    \includegraphics{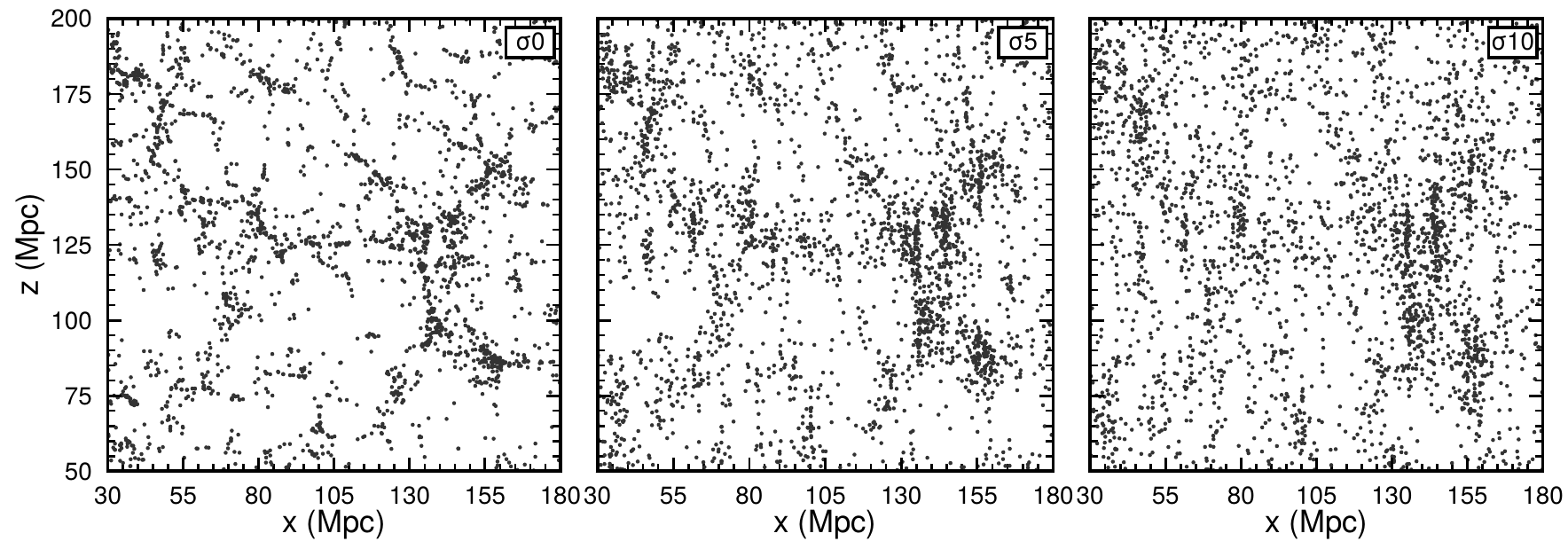}
    \caption{Projection of galaxy distributions of samples \(\sigma0,\ \sigma5, \ \sigma10\) in a slice with a thickness of \SI{10}{Mpc}.
    Each dot represents a galaxy.
    The photometric uncertainties are parallel to the z-axis, which also defines the line of sight in this work.
    Only an area of \SI{150 x 150}{Mpc} is shown for visual clarity.
    For information about samples; see Sect. \ref{subsect:samples}.}
    \label{fig:inputdata}
\end{figure*}

Figure \ref{fig:inputdata} shows the comparison between a sample of galaxies with no uncertainties (all spectroscopic redshifts), and two samples of galaxies with photometric redshifts with uncertainties with distributions \(\mathcal{N}(\SI{0}{Mpc},(\SI{5}{Mpc})^2)\) and \(\mathcal{N}(\SI{0}{Mpc},(\SI{10}{Mpc})^2)\).
The leftmost plot shows a visible web-like structure.
In the middle plot, the structure is more diffuse because of the added randomness along the z-axis, but some of the original structure is still somewhat visible.
In the rightmost plot, the original structure is no longer visible, but rather seems to have filamentary structures along the z-axis that have been produced by the added random errors to galaxy positions along the z-axis.

\subsection{Samples} \label{subsect:samples}

We use the following notation to name the samples.
The fraction of galaxies in the samples with spectroscopic distance estimates is denoted with sXX, where XX is a number indicating the percentage from the whole sample.
The spectroscopic galaxies are always the brightest galaxies in the sample.
For example, s40 means the sample contains 40\% of the brightest galaxies from the whole sample, all of these have exact distances, and is missing the other 60\% of the galaxies.
The photometric samples are denoted with \(\sigma\)YY, where YY is a number indicating the size of the uncertainties for photometric distance estimates.
For example, \(\sigma\)5 means the sample contains galaxies that have uncertainties in the distance measures that are generated with Gaussian distribution with a standard deviation of \SI{5}{Mpc}.
For mixed samples, \(\sigma\)10s30 for example means 30\% of the brightest galaxies have exact distances (i.e. spectroscopic distance estimates), and the rest, that is 70\% of the galaxies in the sample, have distances with uncertainties generated with Gaussian distribution with a standard deviation of \SI{10}{Mpc}.
Table \ref{tab:samples} lists samples used in this work, the distributions used to generate uncertainties for distances, and the percentages of galaxies with spectroscopic distances.

For brevity, hereafter the term spectroscopic galaxies/data is used as a synonym for galaxies/data with spectroscopic redshifts, and photometric galaxies/data is used as a synonym for galaxies/data with photometric redshift.
In this work, the former means data with no uncertainties, and the latter means data with uncertainties along one axis.

\begin{table}[]
    \centering
    \caption{Photometric distance uncertainties and percentage of spectroscopic galaxies in each sample.
    The distance uncertainties column shows the Gaussian distribution used to generate uncertainties for distances of photometric galaxies.
    The last column shows the percentage of the brightest galaxies with spectroscopic distances, i.e. exact distances.
    Samples that do not have galaxies with photometric distances are indicated with an em dash (---) in the second column.}
    \begin{tabular}{c c c}
        \hline\hline
        Name & Distance uncertainties & \multicolumn{1}{p{2cm}}{Spectroscopic\newline distances} \\ \hline
        \(\sigma\)0 & --- & 100\% \\
        \(\sigma\)1 & \(\mathcal{N}(\SI{0}{Mpc}, (\SI{1}{Mpc})^2)\) & 0\% \\
        \(\sigma\)2 & \(\mathcal{N}(\SI{0}{Mpc}, (\SI{2}{Mpc})^2)\) & 0\% \\
        \(\sigma\)3 & \(\mathcal{N}(\SI{0}{Mpc}, (\SI{3}{Mpc})^2)\) & 0\% \\
        \(\sigma\)5 & \(\mathcal{N}(\SI{0}{Mpc}, (\SI{5}{Mpc})^2)\) & 0\% \\
        \(\sigma\)7 & \(\mathcal{N}(\SI{0}{Mpc}, (\SI{7}{Mpc})^2)\) & 0\% \\
        \(\sigma\)10 & \(\mathcal{N}(\SI{0}{Mpc}, (\SI{10}{Mpc})^2)\) & 0\% \\
        \(\sigma\)5s50 & \(\mathcal{N}(\SI{0}{Mpc}, (\SI{5}{Mpc})^2)\) & 50\% \\
        \(\sigma\)5s40 & \(\mathcal{N}(\SI{0}{Mpc}, (\SI{5}{Mpc})^2)\) & 40\% \\
        \(\sigma\)5s30 & \(\mathcal{N}(\SI{0}{Mpc}, (\SI{5}{Mpc})^2)\) & 30\% \\
        \(\sigma\)5s20 & \(\mathcal{N}(\SI{0}{Mpc}, (\SI{5}{Mpc})^2)\) & 20\% \\
        \(\sigma\)5s10 & \(\mathcal{N}(\SI{0}{Mpc}, (\SI{5}{Mpc})^2)\) & 10\% \\
        \(\sigma\)10s50 & \(\mathcal{N}(\SI{0}{Mpc}, (\SI{10}{Mpc})^2)\) & 50\% \\
        \(\sigma\)10s40 & \(\mathcal{N}(\SI{0}{Mpc}, (\SI{10}{Mpc})^2)\) & 40\% \\
        \(\sigma\)10s30 & \(\mathcal{N}(\SI{0}{Mpc}, (\SI{10}{Mpc})^2)\) & 30\% \\
        \(\sigma\)10s20 & \(\mathcal{N}(\SI{0}{Mpc}, (\SI{10}{Mpc})^2)\) & 20\% \\
        \(\sigma\)10s10 & \(\mathcal{N}(\SI{0}{Mpc}, (\SI{10}{Mpc})^2)\) & 10\% \\
        s50 & --- & 50\% \\
        s40 & --- & 40\% \\
        s30 & --- & 30\% \\
        \hline
    \end{tabular}
    \label{tab:samples}
\end{table}

\section{The Bisous filament finder} \label{sect:bisous}

We used the Bisous filament finder to detect the filaments from the mock data.
This finder is a stochastic tool to identify the spines of the filaments using the spatial distribution of galaxies or haloes \citep{Tempel14, Tempel16}.
The Bisous has already been applied to a variety of data and has been proven to give similar results to other filament finders \citep{Libeskind18}.
We give a short overview of the method below.

First, the Bisous randomly populates the volume with points with parameters (called marked points), where each point represents the centre of a cylinder and the parameters give the size and orientation of the cylinder.
The cylinder's width is about \SI{1}{Mpc}, which defines the width of the detected filaments.
This width is derived from the gradient of the galaxy density, where there is a peak at approximately \SI{0.5}{Mpc} from the filament's spine.
Each configuration of cylinders in the volume has a defined energy, which depends on the position of the cylinders in relation to the underlying data of haloes and the interconnectedness of the filamentary network made up of the cylinders.
Using the Metropolis-Hastings algorithm and the simulated annealing procedure, the Bisous model optimises the energy function of the system by suggesting random moves to add, remove, or change the cylinders.

The data of the cylinder configurations are collected over hundreds of thousands of cycles, each consisting of tens of thousands of moves, which is the basis for visit map calculations.
In general, one realisation of cylinders in the volume represents the detected filamentary network.
As the model is stochastic, the configuration of cylinders changes from realisation to realisation.
The combination of many realisations allows us to define the visit map that describes the detected filamentary network.
Each coordinate has a defined visit map value, ranging from 0~to~1.
The visit map contains information on how often a coordinate in space was `visited' by a cylinder, which signifies how probable it is that a random realisation has a cylinder at that position.

To decrease the effects of Poisson noise, the Bisous model is run many times, usually 50-100.
This increases the signal-to-noise ratio as a larger number of independent realisations are combined to obtain the resulting maps.

\citealt{Muru21} show how the galaxy number density affects the detected filamentary network.
These authors show that the Bisous method underestimates the extent of the filamentary network rather than giving false-positive results.
This means that the filament finder underestimates the filamentary structures at higher distances where the galaxy number density drops.
To improve the quality of the detected filamentary network, we need to increase the galaxy number density, for example, with photometric data.

\subsection*{Using photometric data} \label{sect:bisous:photodata}

Filament finders usually need precise data, either scalar or tensor fields or galaxy positions, and therefore the less accurate photometric data are ignored.
Here we present a method that benefits from photometric data by having higher input data density and is able to mitigate the uncertainty from distance measures.
This subsection gives an overview of a simple method of how the Bisous filament finder can use photometric data.

For each galaxy with a photometric redshift estimate and its probability distribution, we generate \(N_R\) new distance estimates {drawn from the photometric redshift probability distribution.}
For the mock data in this paper, we used a Gaussian distribution to generate the uncertainties, and so we also use the same Gaussian distribution to generate different distance estimates for every galaxy.
Every Bisous run uses a different distance estimate for a galaxy with a photometric distance measure.
The number of Bisous runs should be large in order to minimise the Poisson noise in the results but also small to minimise the computational resources used for the model.
Usually, there are around \numrange{50}{100} Bisous runs, for this work, we used \(N_R = 80\), which has been shown to give good results in previous works using the Bisous model.
For mixed data sets of spectroscopic and photometric targets, only the photometric ones have different distance estimates, whereas spectroscopic targets have the same distance value in every run.

The novelty of this method is that the runs that have more accurate distance estimates for the photometric galaxies produce more persistent filaments.
Galaxies with inaccurate distance estimates generate noise.
The Bisous model suppresses the noise by combining a large number of realisations.
The more inaccurate distance estimates there are, the more noise, which means the Bisous is able to find fewer filaments.
This means that uncertainties still have to be small to produce good results.
The generation of new distance estimates is done separately for each galaxy.
In practice, we can use a different probability distribution for each galaxy.

\section{Results} \label{sect:results}

This work uses three types of samples: spectroscopic, photometric, and mixed.
The primary purpose of spectroscopic-only samples is to be a reference value for the other two types of samples.
Photometric-only samples show what can be done using only photometric redshift surveys, and mixed samples show what we can do by combining the spectroscopic and photometric redshift surveys, for example in the areas of spectroscopic surveys where galaxies are sparse or at higher distances where the detection is less complete.

\begin{figure*}
    \centering
    \includegraphics{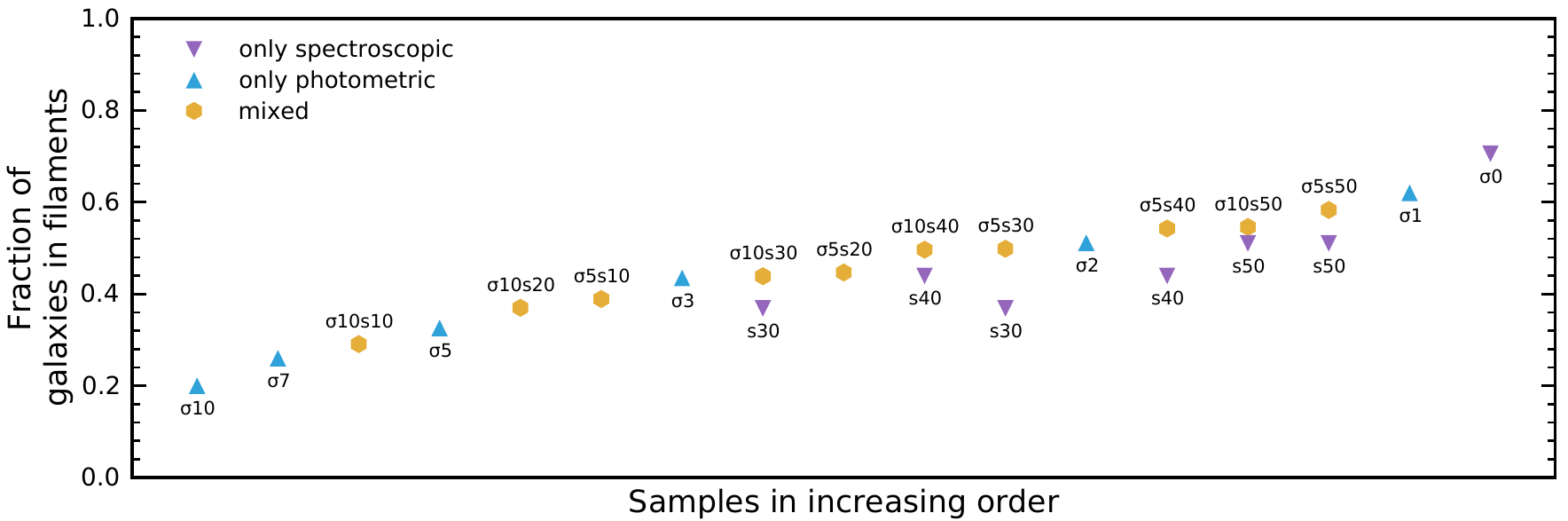}
    \caption{Fraction of galaxies in filaments for different spectroscopic-only, photometric-only, and mixed samples.
    The samples are ordered so that the y-axis values of photometric-only and mixed samples are in increasing order.
    The spectroscopic-only samples are used for reference values to show the increase in the fraction of galaxies in filaments for mixed samples.
    The sample s30 is the smallest spectroscopic sample in this study because smaller samples had too few galaxies to be able to detect the filamentary network.}
    \label{fig:gal_in_fil_all}
\end{figure*}

\begin{figure*}
    \centering
    \includegraphics{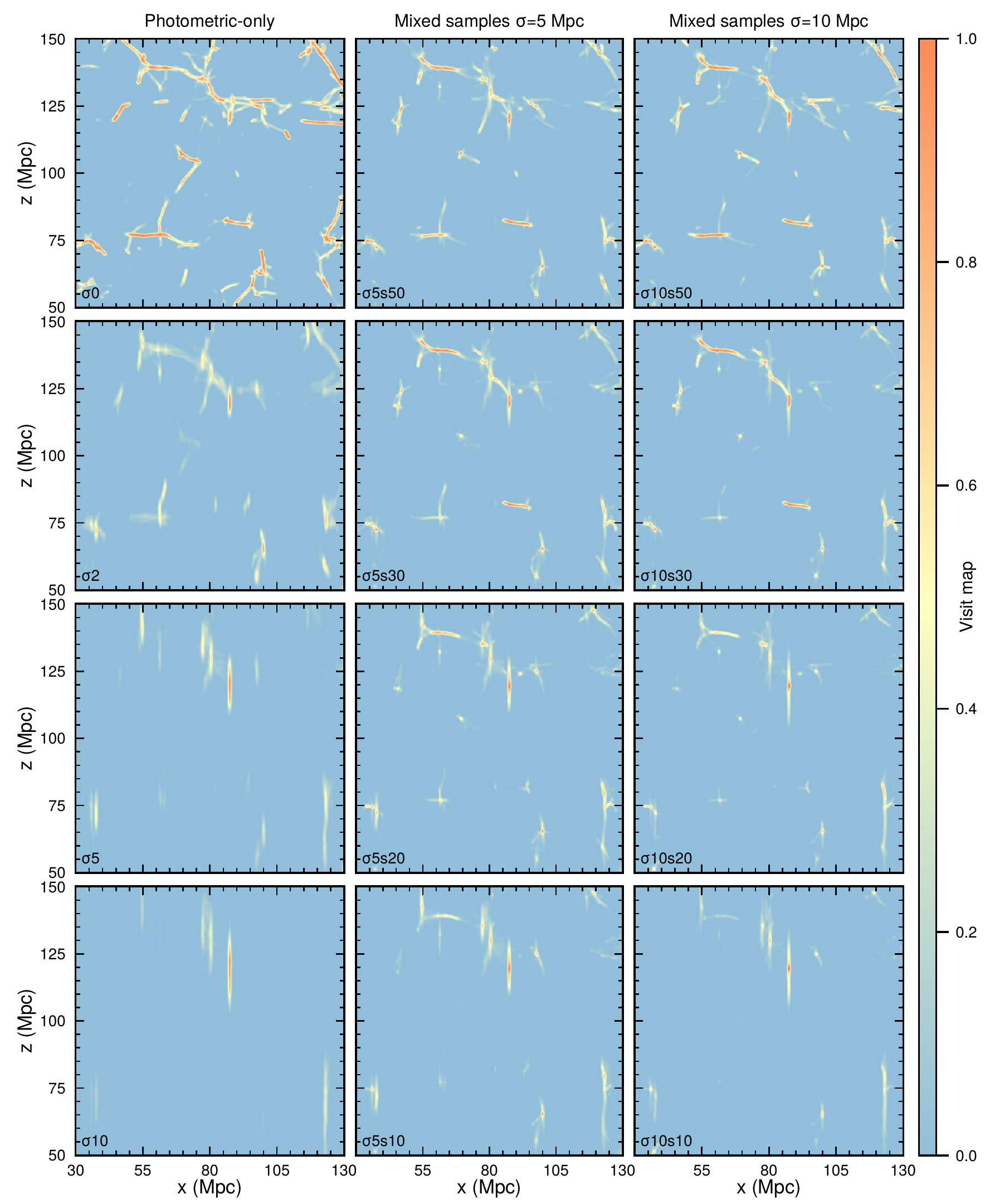}
    \caption{Projections of maximum visit map values in slices obtained from the Bisous model using different samples.
    Only a smaller \SI{100 x 100}{Mpc} area is shown for visual clarity.
    The thickness of the slice is \SI{10}{Mpc}.   
    Usually, a visit map limit of 0.05 is used to classify whether or not a coordinate is inside a filament.
    Therefore, everything besides the blue area is likely part of the filamentary network.
    The vertical axis (z) is parallel to the axis of the photometric uncertainties, i.e. it emulates the line of sight.
    The leftmost column shows samples with only photometric galaxies, the middle column shows mixed samples with medium uncertainties (\(\sigma = \SI{5}{Mpc}\)) for photometric galaxies, and the rightmost column shows mixed samples with the larger uncertainties (\(\sigma = \SI{10}{Mpc}\)).
    Different rows in the leftmost column have different photometric uncertainties, and the middle and the rightmost column have different fractions of the brightest galaxies as spectroscopic galaxies.
 See Table \ref{tab:samples} and Sect. \ref{subsect:samples} for the sample naming convention used here.}
    \label{fig:visitmaps}
\end{figure*}

As mentioned in Section \ref{sect:introduction}, filaments affect the evolution of galaxies and knowing whether a galaxy is in a filament or not is useful when studying the galaxy properties.
Therefore, one of the simplest metrics with which to compare the resulting filamentary network is the fraction of galaxies situated inside filaments.
Figure \ref{fig:gal_in_fil_all} shows the fraction of galaxies inside filaments for all the samples used in this work.
The sample \(\sigma0\) is the most complete sample (galaxy positions without uncertainties but with the same magnitude limit as other samples), and the fraction of galaxies in filaments for that sample could be considered as a reference value for an ideal case.
Looking at samples with only photometric redshift galaxies, we see the expected trend that the larger the uncertainties for the distance, the fewer galaxies are in filaments.
This comes from the fact that the larger the uncertainties, the fewer filaments the Bisous model is able to detect (cf. Fig. \ref{fig:visitmaps}) as the structure in the  galaxy distribution is less obvious, as seen from Figure \ref{fig:inputdata}.
Adding spectroscopic redshift galaxies to create the mixed samples considerably increases the number of galaxies in filaments.
For example, 33\% of the galaxies in \(\sigma5\) are   in filaments, but when 20\% of the brightest galaxies have spectroscopic redshifts (\(\sigma5\)s20) this fraction rises to 45\%, and with 50\% galaxies with spectroscopic redshifts (\(\sigma5\)s50) up to 59\% of galaxies are in filaments.
This shows that using spectroscopic galaxies together with photometric galaxies increasingly improves the detected filamentary network as the number of spectroscopic galaxies in a sample increases.
On the other hand, when comparing spectroscopic-only samples (e.g. s50 or s40) to mixed samples (e.g. \(\sigma5\)s50 or \(\sigma5\)s40) where the galaxy number density is increased with added photometric galaxies, we can see that the mixed samples have more galaxies in filaments when compared to spectroscopic samples.
This indicates that adding galaxies with photometric redshifts to increase the number density of galaxies in the sample helps to improve the detected filamentary network, as it increases the fraction of galaxies in filaments and is closer to the reference sample (\(\sigma0\)).

This metric can also be used to compare the results with observational data, but different filament finders and different filament definitions give results that are not directly comparable.
For example, \citet{Tempel14} found that when using the Bisous model on SDSS data, the fraction of galaxies in filaments is about 40\%, but they use a stricter definition for whether a galaxy is considered to be in a filament or not.
Also, results based on observational data are likely missing fainter galaxies that are present in simulations, which affects the fraction of galaxies in filaments.

It is a good idea to look at the spatial distribution of filaments  produced by different samples to assess them visually.
Figure \ref{fig:visitmaps} shows visit map slices from 12 different samples.
The colour indicates the likelihood of a coordinate being inside a filament.
The plot in the upper left corner is the sample we use as ground truth, the full spectroscopic sample.
The vertical axis is parallel to the axis of photometric uncertainties and emulates the line of sight.
The photometric-only samples in the left column show that photometric galaxies make it very difficult to detect filaments perpendicular to the line of sight.
Only stretched-out filaments parallel to the line of sight remain.
In the middle and rightmost columns, mixed samples are used.
Including the spectroscopic galaxies helps detect filaments perpendicular to the line of sight.
But even in the mixed samples, when photometric galaxies dominate, as in the lower rows, the filaments are preferentially parallel to the line of sight.
This does not mean that filaments are parallel to the line of sight, but that these are the filaments the Bisous model is able to detect with the corresponding data.

\begin{figure*}
    \centering
    \includegraphics{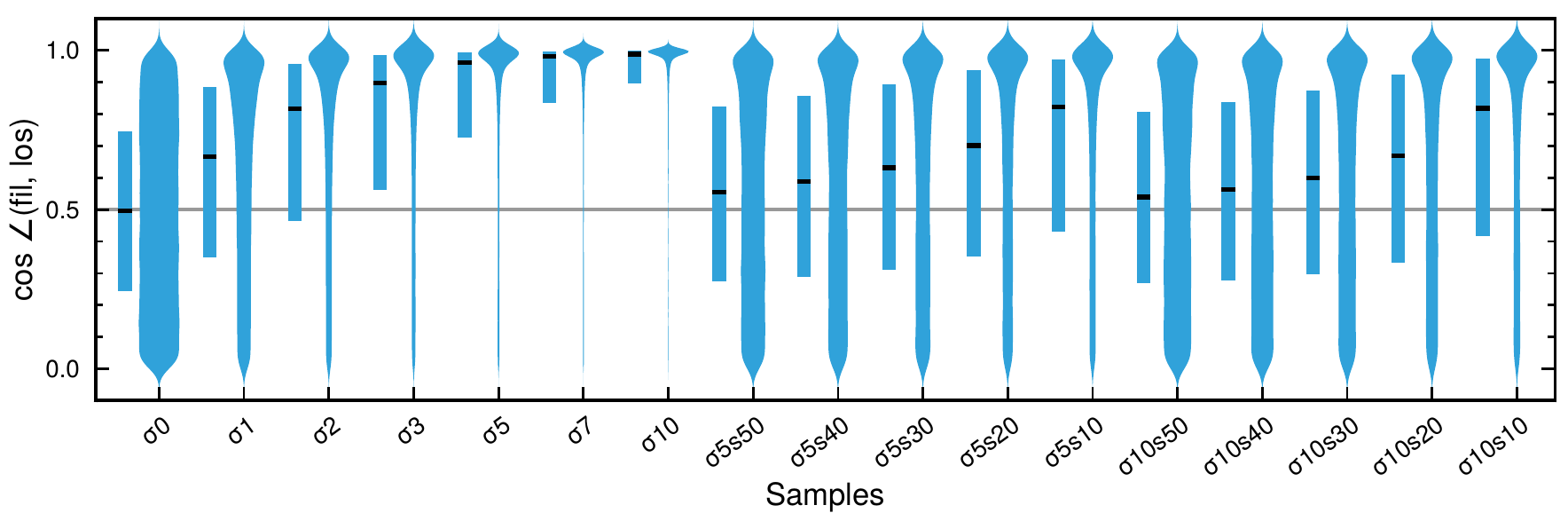}
    \caption{Distributions of the cosine of the angle between filament spines (fil) and the line of sight (los).
    For each sample, there are two plots.
    The left one is a bar plot of the quartiles of the distribution, where the black crossbar indicates the second quartile (the median).
    The right plot is a violin plot that shows the density curve of the distribution.
    The horizontal grey line indicates the median value for a uniform distribution.
    The closer the distribution gets to value 1, the more filaments are parallel to the line of sight (z-axis in other plots).}
    \label{fig:alignment}
\end{figure*}

Figure \ref{fig:visitmaps} shows that photometric galaxies, which have large uncertainties along the line of sight, suppress the detection of filaments perpendicular to the line of sight.
To study this effect, we describe the distribution of angles between filament spines and the line of sight.
These results are shown in Figure \ref{fig:alignment}.
Again, the \(\sigma0\) sample is the baseline for this work and shows a uniform distribution of angles.
Using photometric-only samples skews the distribution closer to 1, meaning the filaments are mostly parallel to the line of sight, as is visible from the visit map projections in Figure \ref{fig:visitmaps}.
Adding spectroscopic galaxies to the samples significantly reduces the bias of high cosine values in the distributions.
This is also visible in Figure \ref{fig:visitmaps}, where more filaments are perpendicular to the z-axis in mixed samples.

\begin{figure*}
    \centering
    \includegraphics[width = 17cm]{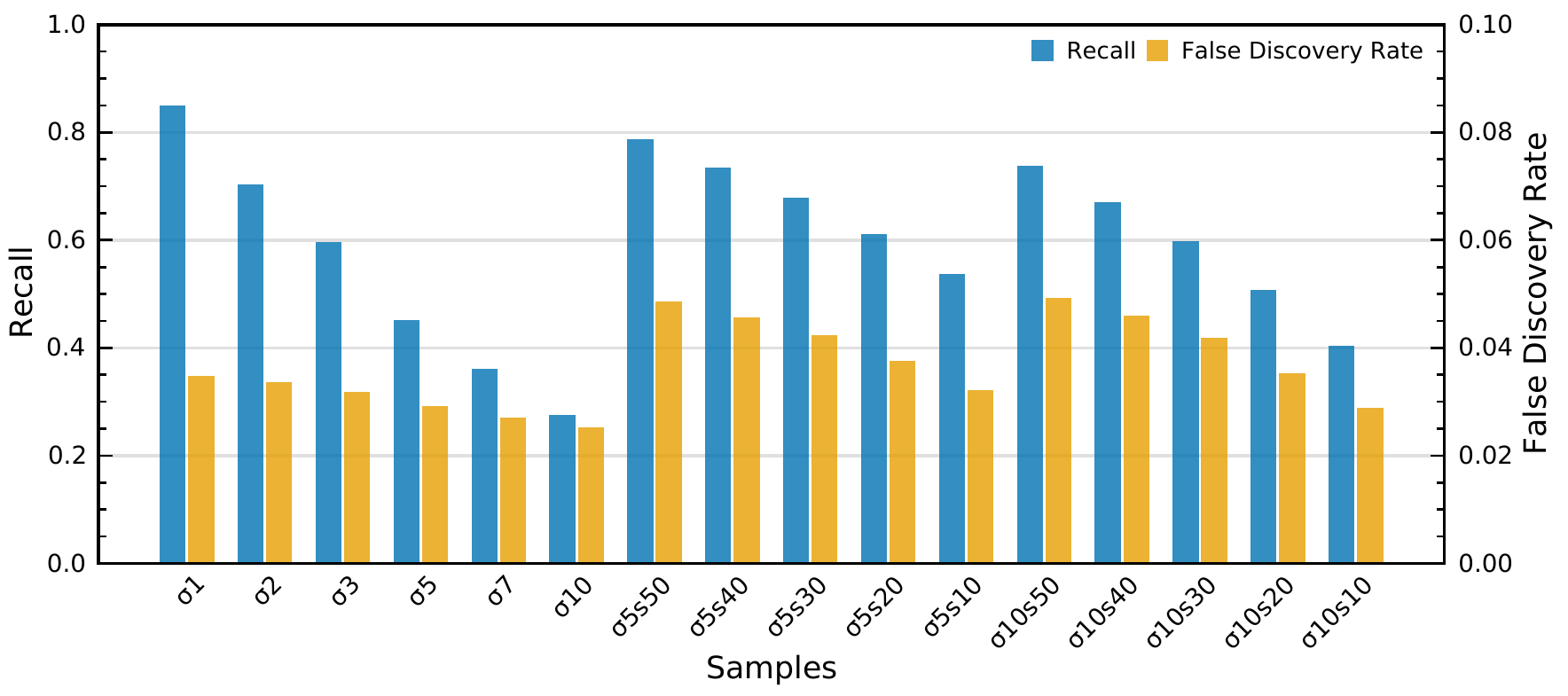}
    \caption{Recall and false discovery rates for photometric and mixed samples.
    All the samples have the same total number of galaxies.
    The definitions for recall and false discovery rate are given in Sect. \ref{sect:results}.
    The false discovery rate uses the secondary vertical axis on the right side of the plot.
    Including spectroscopic galaxies improves recall but also increases false discovery rates.
    The false discovery rates are below 5\% for every sample.}  
    \label{fig:recall_all}
\end{figure*}

When using the results obtained with a full spectroscopic sample \(\sigma0\) as ground truth, we can compare other results to it and construct contingency tables called confusion matrices.
We assign a binary label for each coordinate depending on the visit map value.
If the visit map value is equal to or greater than 0.05, then that coordinate is classified as inside a filament.
With each coordinate labelled, we can assign four kinds of results: true positive, true negative, false positive, and false negative.
To describe the goodness of the results for sample \(s\), we use two statistics: the recall
\begin{equation} \label{eq:recall}
    \text{Recall}_s = \frac{\mathrm{TP}_s}{\mathrm{P}_{\sigma0}} \ ,
\end{equation}
where \(\mathrm{TP}_s\) is the number of true-positive values in the sample \(s\), and \(\mathrm{P}_{\sigma0}\) the number of positive values in the reference sample \(\sigma0\);\\
and the false discovery rate,
\begin{equation} \label{eq:fdr}
    \text{False discovery rate}_s = \frac{\mathrm{FP}_s}{\mathrm{P}_s} \ ,
\end{equation}
where \(\mathrm{FP}_s\) is the number of false-positive values in the sample \(s\), and \(\mathrm{P}_s\) the number of all positive values in the sample \(s\), which includes both the true-positive and false-positive values.
Recall shows the fraction of filaments the model is able to find compared to the filaments present in results obtained with the sample \(\sigma0\), which we want to maximise.
The false discovery rate describes the fraction of false filaments in the results, which we want to minimise.

Figure \ref{fig:recall_all} shows the recall and the false discovery rates for different samples.
As expected, the recall decreases monotonically when photometric uncertainties increase.
Using mixed samples improves the recall even when using small fractions of spectroscopic galaxies.
For example, this improvement can be seen when comparing the recalls of \(\sigma5\) (0.45) and \(\sigma5\)s10 (0.54) or \(\sigma10\) (0.27) and \(\sigma10\)s10 (0.40), both mixed samples use only 10\% of the spectroscopic galaxies.
Using 50\% of the spectroscopic galaxies boosts the recall above 0.73, which means almost three-quarters of the original filaments are detected.
As seen in Figure \ref{fig:recall_all}, the false discovery rate is below 0.05 for every sample.
This shows that the Bisous model produces only little noise and false-positive values even with photometric redshift data.

In addition, we ran Bisous on mock data without using the method described in Sect. \ref{sect:bisous:photodata} and using the samples \(\sigma5\), \(\sigma\)5s30, \(\sigma10\), and \(\sigma\)10s30 as they are.
This enables us to compare the Bisous model results obtained with the method in Sect. \ref{sect:bisous:photodata} with results obtained with photometric data without doing anything special to the photometric galaxies and ignoring photometric redshift errors.
Table \ref{tab:plainruns} lists the different statistics introduced in this section calculated for these Bisous runs.
These results are calculated as a reference and motivation for using the method described in Sect. \ref{sect:bisous:photodata}.
In comparison to the samples introduced in Sect. \ref{subsect:samples} these results show significantly worse recall values and fewer galaxies in filaments.
In some cases, the false discovery rate can have better results, but this comes from the fact that when detecting fewer filaments, there are also fewer false-positive results and therefore a lower false discovery rate.

\begin{table}[]
    \centering
    \caption{Comparison of Bisous model results with and without using the method described in Sect. \ref{sect:bisous:photodata}.
    PB in front of the sample name indicates that the results are obtained with the plain Bisous model.
    Recall and false discovery rate are defined by Equations \ref{eq:recall} and \ref{eq:fdr}.}
    \begin{tabular}{l c c c}
        \hline \hline
        Sample & gal in fil\(^a\) & Recall & FDR\(^b\) \\ \hline
        PB \(\sigma5\) & 0.217 & 0.296 & 0.037 \\ \vspace{4pt}
        \(\sigma5\) & 0.329 & 0.452 & 0.029 \\
        PB \(\sigma5\)s30 & 0.398 & 0.541 & 0.039 \\ \vspace{4pt}
        \(\sigma5\)s30 & 0.501 & 0.679 & 0.042 \\
        PB \(\sigma10\) & 0.127 & 0.172 & 0.039 \\ \vspace{4pt}
        \(\sigma10\) & 0.200 & 0.276 & 0.025 \\
        PB \(\sigma10\)s30 & 0.348 & 0.473 & 0.039 \\
        \(\sigma10\)s30 & 0.441 & 0.598 & 0.042 \\ \hline
    \end{tabular}
    \tablefoot{\(^a\) Fraction of galaxies in filaments; \(^b\) false discovery rate}
    \label{tab:plainruns}
\end{table}

These results qualitatively confirm the results from \citet{Kruuse2019}, showing that galaxies with photometric redshifts are clustered around the Bisous filaments.
We show that the Bisous model can use photometric redshift data to detect the filamentary network without producing significant amounts of false-positive results.
However, when the uncertainties in the distance measure increase, the model is able to recall fewer filaments.
For example, with sample \(\sigma10\), the recall is only 0.27, and mostly filaments parallel to the line of sight are detected.
Including spectroscopic galaxies in the samples considerably improves the recall and helps to mitigate the issue with filament alignment in the detected filamentary network.
Results in this work also qualitatively follow the results of \citet{Muru21}, where they show how the Bisous filaments depend on the number density of the galaxies in the input data.
In this work, the mixed samples show a similar trend, and the photometric galaxies boost the number density of galaxies, although less than the same number of spectroscopic galaxies would.

\section{Discussion} \label{sect:discussion}

Previous works applied the Bisous model to SDSS, which is a spectroscopic survey, and compiled a catalogue of filaments \citep{Tempel14}.
This work extends the applicability of the model and demonstrates the effects of using data with photometric redshifts.
The major benefit of using photometric redshift data comes from its comparatively high availability, and measurements are made in bulk, and not for single galaxies as in spectroscopic measurements.
The problem with photometric redshift data is the significantly larger uncertainties when calculating redshifts, which results in larger uncertainties in distance measurements.
This is problematic for filament finders.

To simulate the large uncertainties in distance measurements, we used simulation data to create data with added uncertainties.
For simplicity, all the uncertainties are generated with the same Gaussian distribution for each galaxy.
In reality, the uncertainty depends on many properties, and one of the most relevant is the magnitude of the galaxy.
But the dependence of the uncertainty on the magnitude is different for different surveys.
Also, using a simulation removes any redshift dependence in the data.
In observations, there are two major redshift-dependent effects.
Firstly, the number density of galaxies decreases with redshift as we are able to detect fewer galaxies on the fainter end, and this affects the ability of the  Bisous
model to detect filaments as shown in \citet{Muru21}.
Secondly, the precision of the photometric redshift values for galaxies depends on their actual distance.
These dependencies should be studied in greater depth when concentrating on specific surveys and is outside the scope of this study.

The method we use to overcome this problem of large uncertainties is straightforward.
Essentially, we are just guessing the true positions.
Each galaxy gets 80 different random positions based on the uncertainties of the redshift estimate.
The theory behind this approach is that while random inaccurate positions produce noise, the positions close to the true position of the galaxy produce a strong enough signal to be above the noise level.
Regardless of its simplicity, the method shows considerable improvements over results when not using this method (see Table \ref{tab:plainruns}).

Although this simple method improves the results, the problems introduced by using the photometric redshift data are still prevalent.
Using photometric-only redshift data (\(\sigma\)XX samples) results in part of the signal being lost and an incomplete filamentary network.
This is visible from the recall values when compared against \(\sigma0\) (Fig. \ref{fig:recall_all}), the fraction of galaxies in filaments (Fig. \ref{fig:gal_in_fil_all}), and the projections of visit map values (Fig. \ref{fig:visitmaps}).
Another problem is that with larger uncertainties for distances, the filaments perpendicular to the line of sight are almost impossible to detect. This creates a strong bias for filaments parallel to the line of sight (cf Fig. \ref{fig:alignment}).
It is important to note that the false discovery rate (cf Fig. \ref{fig:recall_all}) decreases when data with larger uncertainties are used.
This is because galaxies with larger uncertainties produce less meaningful signals, and therefore there will be fewer filaments in the results, which also means fewer false-positive filaments.
Low false discovery rate values are good because they demonstrate the robustness of the results.
The model rather outputs fewer filaments than false-positive filaments.

All of the aforementioned problems are reduced by using mixed samples of spectroscopic and photometric redshift data instead only photometric, as shown in Section \ref{sect:results}.
Figure \ref{fig:gal_in_fil_all} also shows that using mixed samples to boost the galaxy number density is better than only using the spectroscopic redshift galaxies.
This could be useful, for example, in the more distant areas of spectroscopic surveys, where galaxies with spectroscopic redshifts are too sparse to use for the detection of the large-scale structure.
Using mixed data could help us extend the area where we can reliably detect the filaments.

Still, this method requires photometric redshift data with relatively small uncertainties, which are not usually achieved by photometric surveys.
Unfortunately, all current photometric surveys have unusably large uncertainties for the redshifts, but there will be some new surveys with suitable accuracy in the near future.
One prominent candidate for photometric redshift data is the upcoming Javalambre Physics of the Accelerating Universe Astrophysical Survey (J-PAS, \citealt{JPAS, miniJPAS, Laur2022}).
J-PAS is designed to measure the positions and redshifts of 14 million galaxies.
And the estimated precision for the photometric redshift for galaxies in the redshift range \(0.1 < z < 1.2\) is \(\sigma_z \lesssim 0.003(1+z)\).
For example, when using SDSS, the spectroscopic redshift galaxy number density is high enough to detect some filaments up to a distance of \SI{400}{Mpc}, which is approximately \(z = 0.1\) \citep{Tempel14, Muru21}.
For this distance, the precision of the redshifts is \(\sigma_z \lesssim 0.003 \times 1.1 \approx \SI{14}{Mpc}\).
This is the same order of magnitude as the \(\sigma10\) samples used in this work.
We expect the uncertainties to be smaller for brighter galaxies.
We aim to apply the Bisous model to J-PAS data when it is released and compile a catalogue of filaments.
To obtain the mixed data of photometric and spectroscopic redshift galaxies, we plan to use the Sloan Digital Sky Survey (SDSS) \citep{SDSS_DR12} and the Dark Energy Spectroscopic Instrument (DESI) Bright Galaxy Survey (BGS) \citep{DESI, DESI_BGS}.

Although this study is based on the Bisous filament finder, it is likely that the general tendencies when using data with photometric redshifts are similar with other filament finders.
Using photometric data will decrease the effectiveness of the filament finder, and filaments parallel to the line of sight are more likely to be detected.
It is uncertain whether using mixed data of photometric and spectroscopic redshifts improves the results compared to using only spectroscopic data when using other filament finders.
Also, the false discovery rates might have different values for other filament finders.
The advantage of the Bisous model is that it models the underlying filamentary network, and galaxies are only used to constrain the model properties.
Hence, in the Bisous filament finder, it is straightforward to combine spectroscopic and photometric samples.
While fixing the scale of the filaments in the Bisous model, we are free from smoothing the galaxy distribution, and the Bisous model is able to detect filaments with a specified scale regardless of the galaxy density.

As mentioned in Sect. \ref{sect:introduction}, one common application for filaments is to study the alignment of galaxies and their host filaments.
This means that obtaining the accurate filament orientation from the data is instrumental.
In future studies, we aim to improve the Bisous model to reduce the alignment bias of filaments when using data with photometric redshifts.

\section{Conclusions} \label{sect:conclusions}

Filament finders are limited, among other things, by the abundance of spectroscopic redshift data.
This limits the sky areas and depth where we can detect the filamentary network.
As photometric redshift data can be obtained on shorter timescales, because you can observe many objects simultaneously, there are  many more photometric redshift data available.
We present a method that enables the Bisous filament finder to use data with considerable uncertainties in one coordinate; for example photometric redshift data.
We use \textsc{MultiDark-Galaxies}, a dark matter-only simulation with semi-analytical galaxies, to generate the data for analysis.
Spectroscopic redshift data are simply the exact positions of galaxies from the simulation, and photometric redshift galaxies have added random error with Gaussian distribution in one axis, where this latter represents the line of sight.
This work uses three types of samples.
Firstly, spectroscopic samples with different magnitude cuts for reference values for other samples.
Secondly, photometric samples using different standard deviations from \(\sigma = \) \SIlist[list-pair-separator = { to }]{1;10}{Mpc} to generate the errors with different sizes for distances.
Thirdly, mixed samples, where in different samples \SIlist[list-pair-separator = { to }]{10;50}{\%} of the brightest galaxies have spectroscopic redshifts, that is, they have exact distance measurements, and other galaxies have distances with uncertainties.
An overview of the samples used in this work is given in Sect. \ref{subsect:samples}.

The Bisous model uses a marked point process to fit cylinder-like objects to the underlying galaxy distribution and optimises the distribution of objects based on the galaxy distribution and the interconnectedness of the cylinder network.
To use the photometric redshift data with uncertainties along one axis, we modified the coordinates along that axis.
Knowing the distribution of the uncertainties for the distance of photometric redshift galaxies, we use the same distribution to add a random value to the distance of a galaxy.
For each Bisous run, we generated a new galaxy distribution, where each photometric redshift galaxy has a different random value added to its distance based on the uncertainty distribution.
Each Bisous model uses 80 runs.
The theory underpinning this approach is that those runs, where some galaxies have random distance values that are closer to their true distances, produce strong signals, and others with scrambled galaxy distributions just produce noise, which will be removed in the post-processing.

Using photometric-only samples shows that when uncertainties are very small, Gaussian distribution with  \(\sigma = \) \SI{1}{Mpc} or \SI{2}{Mpc}, the Bisous model can find most of the same filaments as in the full spectroscopic sample \(\sigma0\).
Unfortunately, these uncertainties are unachievable for modern or even future planned photometric surveys.
With larger uncertainties in the photometric-only samples, the ability to recall the filaments in the reference sample drops below 50\%, and the filaments align with the line of sight.
Using mixed samples of photometric and spectroscopic data helps to reduce the mentioned problems.
For example, a comparison between three samples:
a spectroscopic-only sample s30, which uses only 30\% of the brightest galaxies,
a photometric-only sample \(\sigma\)10, which uses data with errors generated with \(\sigma = \SI{10}{Mpc}\),
and a mixed sample \(\sigma\)10s30, which uses the same standard deviation (\(\sigma = \SI{10}{Mpc}\)) for errors and the same amount of spectroscopic galaxies (30\% of the whole sample).
Using the spectroscopic data, which contain only 30\% of the brightest galaxies, results in 36\% of galaxies being inside filaments.
Using only photometric data, which contain all the galaxies, but have uncertainties in one coordinate, we find that 20\% of galaxies are inside filaments.
And finally, using the mixed data, which contain more data than the spectroscopic sample and, in contrast to the photometric sample, also incorporate 30\% of the spectroscopic data, the Bisous model finds that 40\% of the galaxies are inside filaments.
The reference value for these galaxies and the volume is from the full spectroscopic sample, which gives a value of 71\% of galaxies in filaments.
Adding the spectroscopic galaxies from the sample s30 to the photometric sample \(\sigma\)10 increases the recall of filaments from 27\% to 60\%.
This shows that using mixed data is beneficial when spectroscopic data are too sparse and photometric data have excessively large uncertainties to be used without spectroscopic data.

J-PAS is an upcoming photometric survey that is designed to produce data with sufficiently small uncertainties to be applicable to a method such as the one in this article.
The expected precision of the redshifts is \(\sigma_z \lesssim 0.003(1+z)\) \citep{JPAS}.
For a distance of about \(z = 0.1\), this is \(\sigma_z \lesssim 0.003 \times 1.1 \approx \SI{14}{Mpc}\), which is close to the values used in this work.
The next step is to apply the Bisous model to J-PAS data once available.

\begin{acknowledgements}
We thank the referee for their comments and suggested improvements.

Part of this work was supported by institutional research funding PRG1006 of the Estonian Ministry of Education and Research. We acknowledge the support by the Centre of Excellence "Dark Side of the Universe" (TK133).

Part of this work was carried out in the High-Performance Computing Center of the University of Tartu \citep{ut_hpc}.

The CosmoSim database used in this paper is a service by the Leibniz-Institute for  Astrophysics  Potsdam  (AIP). The \textsc{MultiDark} database was developed in cooperation with the Spanish MultiDark Consolider Project CSD2009-00064. The authors gratefully acknowledge the Gauss Centre for Supercomputing e.V. (www.gauss-centre.eu) and the Partnership for Advanced Supercomputing in Europe (PRACE, www.prace-ri.eu) for funding the \textsc{MultiDark} simulation project by providing computing time on the GCS Supercomputer SuperMUC at Leibniz Supercomputing Centre (LRZ, www.lrz.de).

The data exploration was done using TOPCAT \citep{TOPCAT}, and analysis and plotting were done using Julia Language \citep{julialang} and the following packages:
DrWatson.jl \citep{julia_drwatson},
Pluto.jl \citep{julia_pluto},
Makie.jl \citep{julia_makie},
DataFrames.jl \citep{julia_dataframes},
Distributions.jl \citep{julia_distributions},
ColorSchemes.jl, which uses Scientific colour maps \citep{colormaps}.
\end{acknowledgements}

\bibliographystyle{aa}
\bibliography{references}

\begin{thebibliography}{41}
\expandafter\ifx\csname natexlab\endcsname\relax\def\natexlab#1{#1}\fi

\bibitem[{{Alam} {et~al.}(2015){Alam}, {Albareti}, {Allende Prieto}, {Anders},
  {Anderson}, {Anderton}, {Andrews}, {Armengaud}, {Aubourg}, {Bailey}, \&
  et~al.}]{SDSS_DR12}
{Alam}, S., {Albareti}, F.~D., {Allende Prieto}, C., {et~al.} 2015, \apjs, 219,
  12

\bibitem[{{Alpaslan} {et~al.}(2015){Alpaslan}, {Driver}, {Robotham},
  {Obreschkow}, {Andrae}, {Cluver}, {Kelvin}, {Lange}, {Owers}, {Taylor},
  {Andrews}, {Bamford}, {Bland-Hawthorn}, {Brough}, {Brown}, {Colless},
  {Davies}, {Eardley}, {Grootes}, {Hopkins}, {Kennedy}, {Liske},
  {Lara-L{\'o}pez}, {L{\'o}pez-S{\'a}nchez}, {Loveday}, {Madore}, {Mahajan},
  {Meyer}, {Moffett}, {Norberg}, {Penny}, {Pimbblet}, {Popescu}, {Seibert}, \&
  {Tuffs}}]{Alpaslan2015}
{Alpaslan}, M., {Driver}, S., {Robotham}, A. S.~G., {et~al.} 2015, \mnras, 451,
  3249

\bibitem[{{Arag{\'o}n-Calvo} {et~al.}(2007){Arag{\'o}n-Calvo}, {van de
  Weygaert}, {Jones}, \& {van der Hulst}}]{AragonCalvo2007}
{Arag{\'o}n-Calvo}, M.~A., {van de Weygaert}, R., {Jones}, B. J.~T., \& {van
  der Hulst}, J.~M. 2007, \apjl, 655, L5

\bibitem[{{Beck} {et~al.}(2016){Beck}, {Dobos}, {Budav{\'a}ri}, {Szalay}, \&
  {Csabai}}]{Beck2016}
{Beck}, R., {Dobos}, L., {Budav{\'a}ri}, T., {Szalay}, A.~S., \& {Csabai}, I.
  2016, \mnras, 460, 1371

\bibitem[{{Benitez} {et~al.}(2014){Benitez}, {Dupke}, {Moles}, {Sodre},
  {Cenarro}, {Marin-Franch}, {Taylor}, {Cristobal}, {Fernandez-Soto}, {Mendes
  de Oliveira}, {Cepa-Nogue}, {Abramo}, {Alcaniz}, {Overzier},
  {Hernandez-Monteagudo}, {Alfaro}, {Kanaan}, {Carvano}, {Reis}, {Martinez
  Gonzalez}, {Ascaso}, {Ballesteros}, {Xavier}, {Varela}, {Ederoclite},
  {Vazquez Ramio}, {Broadhurst}, {Cypriano}, {Angulo}, {Diego}, {Zandivarez},
  {Diaz}, {Melchior}, {Umetsu}, {Spinelli}, {Zitrin}, {Coe}, {Yepes}, {Vielva},
  {Sahni}, {Marcos-Caballero}, {Shu Kitaura}, {Maroto}, {Masip}, {Tsujikawa},
  {Carneiro}, {Gonzalez Nuevo}, {Carvalho}, {Reboucas}, {Carvalho}, {Abdalla},
  {Bernui}, {Pigozzo}, {Ferreira}, {Chandrachani Devi}, {Bengaly}, {Campista},
  {Amorim}, {Asari}, {Bongiovanni}, {Bonoli}, {Bruzual}, {Cardiel}, {Cava},
  {Cid Fernandes}, {Coelho}, {Cortesi}, {Delgado}, {Diaz Garcia}, {Espinosa},
  {Galliano}, {Gonzalez-Serrano}, {Falcon-Barroso}, {Fritz}, {Fernandes},
  {Gorgas}, {Hoyos}, {Jimenez-Teja}, {Lopez-Aguerri}, {Lopez-San Juan},
  {Mateus}, {Molino}, {Novais}, {OMill}, {Oteo}, {Perez-Gonzalez}, {Poggianti},
  {Proctor}, {Ricciardelli}, {Sanchez-Blazquez}, {Storchi-Bergmann}, {Telles},
  {Schoennell}, {Trujillo}, {Vazdekis}, {Viironen}, {Daflon},
  {Aparicio-Villegas}, {Rocha}, {Ribeiro}, {Borges}, {Martins}, {Marcolino},
  {Martinez-Delgado}, {Perez-Torres}, {Siffert}, {Calvao}, {Sako}, {Kessler},
  {Alvarez-Cand al}, {De Pra}, {Roig}, {Lazzaro}, {Gorosabel}, {Lopes de
  Oliveira}, {Lima-Neto}, {Irwin}, {Liu}, {Alvarez}, {Balmes}, {Chueca},
  {Costa-Duarte}, {da Costa}, {Dantas}, {Diaz}, {Fabregat}, {Ferrari},
  {Gavela}, {Gracia}, {Gruel}, {Gutierrez}, {Guzman}, {Hernandez-Fernand ez},
  {Herranz}, {Hurtado-Gil}, {Jablonsky}, {Laporte}, {Le Tiran}, {Licandro},
  {Lima}, {Martin}, {Martinez}, {Montero}, {Penteado}, {Pereira}, {Peris},
  {Quilis}, {Sanchez-Portal}, {Soja}, {Solano}, {Torra}, \&
  {Valdivielso}}]{JPAS}
{Benitez}, N., {Dupke}, R., {Moles}, M., {et~al.} 2014, arXiv e-prints,
  arXiv:1403.5237

\bibitem[{Besançon {et~al.}(2021)Besançon, Papamarkou, Anthoff, Arslan,
  Byrne, Lin, \& Pearson}]{julia_distributions}
Besançon, M., Papamarkou, T., Anthoff, D., {et~al.} 2021, Journal of
  Statistical Software, 98, 1

\bibitem[{Bezanson {et~al.}(2017)Bezanson, Edelman, Karpinski, \&
  Shah}]{julialang}
Bezanson, J., Edelman, A., Karpinski, S., \& Shah, V.~B. 2017, SIAM Review, 59,
  65

\bibitem[{{Bonoli} {et~al.}(2020){Bonoli}, {Mar{\'\i}n-Franch}, {Varela},
  {V{\'a}zquez Rami{\'o}}, {Abramo}, {Cenarro}, {Dupke}, {V{\'\i}lchez},
  {Crist{\'o}bal-Hornillos}, {Gonz{\'a}lez Delgado},
  {Hern{\'a}ndez-Monteagudo}, {L{\'o}pez-Sanjuan}, {Muniesa}, {Civera},
  {Ederoclite}, {Hern{\'a}n-Caballero}, {Marra}, {Baqui}, {Cortesi},
  {Cypriano}, {Daflon}, {de Amorim}, {D{\'\i}az-Garc{\'\i}a}, {Diego},
  {Mart{\'\i}nez-Solaeche}, {P{\'e}rez}, {Placco}, {Prada}, {Queiroz},
  {Alcaniz}, {Alvarez-Candal}, {Cepa}, {Maroto}, {Roig}, {Siffert}, {Taylor},
  {Benitez}, {Moles}, {Sodr{\'e}}, {Carneiro}, {Mendes de Oliveira}, {Abdalla},
  {Angulo}, {Aparicio Resco}, {Balaguera-Antol{\'\i}nez}, {Ballesteros},
  {Brito-Silva}, {Broadhurst}, {Carrasco}, {Castro}, {Cid Fernandes}, {Coelho},
  {de Melo}, {Doubrawa}, {Fernandez-Soto}, {Ferrari}, {Finoguenov},
  {Garc{\'\i}a-Benito}, {Iglesias-P{\'a}ramo}, {Jim{\'e}nez-Teja}, {Kitaura},
  {Laur}, {Lopes}, {Lucatelli}, {Mart{\'\i}nez}, {Maturi}, {Quartin},
  {Pigozzo}, {Rodr{\`\i}guez-Mart{\`\i}n}, {Salzano}, {Tamm}, {Tempel},
  {Umetsu}, {Valdivielso}, {von Marttens}, {Zitrin}, {D{\'\i}az-Mart{\'\i}n},
  {L{\'o}pez-Alegre}, {L{\'o}pez-Sainz}, {Yanes-D{\'\i}az}, {Rueda-Teruel},
  {Rueda-Teruel}, {Abril Iba{\~n}ez}, {Ant{\'o}n Bravo}, {Bello Ferrer},
  {Bielsa}, {Casino}, {Castillo}, {Chueca}, {Cuesta}, {Garzar{\'a}n Calderaro},
  {Iglesias-Marzoa}, {{\'I}niguez}, {Lamadrid Gutierrez}, {Lopez-Martinez},
  {Lozano-P{\'e}rez}, {Ma{\'\i}cas Sacrist{\'a}n}, {Molina-Ib{\'a}{\~n}ez},
  {Moreno-Signes}, {Rodr{\'\i}guez Llano}, {Royo Navarro}, {Tilve Rua},
  {Andrade}, {Alfaro}, {Akras}, {Arnalte-Mur}, {Ascaso}, {Barbosa},
  {Beltr{\'a}n Jim{\'e}nez}, {Benetti}, {Bengaly}, {Bernui}, {Blanco-Pillado},
  {Borges Fernandes}, {Bregman}, {Bruzual}, {Calderone}, {Carvano}, {Casarini},
  {Chies-Santos}, {Coutinho de Carvalho}, {Dimauro}, {Duarte Puertas},
  {Figueruelo}, {Gonz{\'a}lez-Serrano}, {Guerrero}, {Gurung-L{\'o}pez},
  {Herranz}, {Huertas-Company}, {Irwin}, {Izquierdo-Villalba}, {Kanaan},
  {Kehrig}, {Kirkpatrick}, {Lim}, {Lopes}, {Lopes de Oliveira},
  {Marcos-Caballero}, {Mart{\'\i}nez-Delgado}, {Mart{\'\i}nez-Gonz{\'a}lez},
  {Mart{\'\i}nez-Somonte}, {Oliveira}, {Orsi}, {Overzier}, {Penna-Lima},
  {Reis}, {Spinoso}, {Tsujikawa}, {Vielva}, {Vitorelli}, {Xia}, {Yuan},
  {Arroyo-Polonio}, {Dantas}, {Galarza}, {Gon{\c{c}}alves}, {Gon{\c{c}}alves},
  {Gonzalez}, {Gonzalez}, {Greisel}, {Land im}, {Lazzaro}, {Magris},
  {Monteiro-Oliveira}, {Pereira}, {Rebou{\c{c}}as}, {Rodriguez-Espinosa},
  {Santos da Costa}, \& {Telles}}]{miniJPAS}
{Bonoli}, S., {Mar{\'\i}n-Franch}, A., {Varela}, J., {et~al.} 2020, arXiv
  e-prints, arXiv:2007.01910

\bibitem[{{Cautun} {et~al.}(2013){Cautun}, {van de Weygaert}, \&
  {Jones}}]{Cautun2013}
{Cautun}, M., {van de Weygaert}, R., \& {Jones}, B. J.~T. 2013, \mnras, 429,
  1286

\bibitem[{Crameri(2021)}]{colormaps}
Crameri, F. 2021, Scientific colour maps, {The development of the Scientific
  colour maps is not funded any longer, but will continue as a pro bono project
  for the scientific community. - Fabio}

\bibitem[{Danisch \& Krumbiegel(2021)}]{julia_makie}
Danisch, S. \& Krumbiegel, J. 2021, Journal of Open Source Software, 6, 3349

\bibitem[{Datseris {et~al.}(2020)Datseris, Isensee, Pech, \&
  Gál}]{julia_drwatson}
Datseris, G., Isensee, J., Pech, S., \& Gál, T. 2020, Journal of Open Source
  Software, 5, 2673

\bibitem[{{de Jong} {et~al.}(2019){de Jong}, {Agertz}, {Berbel}, {Aird},
  {Alexander}, {Amarsi}, {Anders}, {Andrae}, {Ansarinejad}, {Ansorge},
  {Antilogus}, {Anwand-Heerwart}, {Arentsen}, {Arnadottir}, {Asplund}, {Auger},
  {Azais}, {Baade}, {Baker}, {Baker}, {Balbinot}, {Baldry}, {Banerji},
  {Barden}, {Barklem}, {Barth{\'e}l{\'e}my-Mazot}, {Battistini}, {Bauer},
  {Bell}, {Bellido-Tirado}, {Bellstedt}, {Belokurov}, {Bensby}, {Bergemann},
  {Bestenlehner}, {Bielby}, {Bilicki}, {Blake}, {Bland-Hawthorn}, {Boeche},
  {Boland}, {Boller}, {Bongard}, {Bongiorno}, {Bonifacio}, {Boudon}, {Brooks},
  {Brown}, {Brown}, {Br{\"u}ggen}, {Brynnel}, {Brzeski}, {Buchert},
  {Buschkamp}, {Caffau}, {Caillier}, {Carrick}, {Casagrande}, {Case}, {Casey},
  {Cesarini}, {Cescutti}, {Chapuis}, {Chiappini}, {Childress}, {Christlieb},
  {Church}, {Cioni}, {Cluver}, {Colless}, {Collett}, {Comparat}, {Cooper},
  {Couch}, {Courbin}, {Croom}, {Croton}, {Daguis{\'e}}, {Dalton}, {Davies},
  {Davis}, {de Laverny}, {Deason}, {Dionies}, {Disseau}, {Doel}, {D{\"o}scher},
  {Driver}, {Dwelly}, {Eckert}, {Edge}, {Edvardsson}, {Youssoufi}, {Elhaddad},
  {Enke}, {Erfanianfar}, {Farrell}, {Fechner}, {Feiz}, {Feltzing}, {Ferreras},
  {Feuerstein}, {Feuillet}, {Finoguenov}, {Ford}, {Fotopoulou}, {Fouesneau},
  {Frenk}, {Frey}, {Gaessler}, {Geier}, {Gentile Fusillo}, {Gerhard},
  {Giannantonio}, {Giannone}, {Gibson}, {Gillingham},
  {Gonz{\'a}lez-Fern{\'a}ndez}, {Gonzalez-Solares}, {Gottloeber}, {Gould},
  {Grebel}, {Gueguen}, {Guiglion}, {Haehnelt}, {Hahn}, {Hansen}, {Hartman},
  {Hauptner}, {Hawkins}, {Haynes}, {Haynes}, {Heiter}, {Helmi}, {Aguayo},
  {Hewett}, {Hinton}, {Hobbs}, {Hoenig}, {Hofman}, {Hook}, {Hopgood},
  {Hopkins}, {Hourihane}, {Howes}, {Howlett}, {Huet}, {Irwin}, {Iwert},
  {Jablonka}, {Jahn}, {Jahnke}, {Jarno}, {Jin}, {Jofre}, {Johl}, {Jones},
  {J{\"o}nsson}, {Jordan}, {Karovicova}, {Khalatyan}, {Kelz}, {Kennicutt},
  {King}, {Kitaura}, {Klar}, {Klauser}, {Kneib}, {Koch}, {Koposov},
  {Kordopatis}, {Korn}, {Kosmalski}, {Kotak}, {Kovalev}, {Kreckel}, {Kripak},
  {Krumpe}, {Kuijken}, {Kunder}, {Kushniruk}, {Lam}, {Lamer}, {Laurent},
  {Lawrence}, {Lehmitz}, {Lemasle}, {Lewis}, {Li}, {Lidman}, {Lind}, {Liske},
  {Lizon}, {Loveday}, {Ludwig}, {McDermid}, {Maguire}, {Mainieri}, {Mali},
  {Mandel}, {Mandel}, {Mannering}, {Martell}, {Martinez Delgado}, {Matijevic},
  {McGregor}, {McMahon}, {McMillan}, {Mena}, {Merloni}, {Meyer}, {Michel},
  {Micheva}, {Migniau}, {Minchev}, {Monari}, {Muller}, {Murphy},
  {Muthukrishna}, {Nandra}, {Navarro}, {Ness}, {Nichani}, {Nichol}, {Nicklas},
  {Niederhofer}, {Norberg}, {Obreschkow}, {Oliver}, {Owers}, {Pai},
  {Pankratow}, {Parkinson}, {Paschke}, {Paterson}, {Pecontal}, {Parry},
  {Phillips}, {Pillepich}, {Pinard}, {Pirard}, {Piskunov}, {Plank},
  {Pl{\"u}schke}, {Pons}, {Popesso}, {Power}, {Pragt}, {Pramskiy}, {Pryer},
  {Quattri}, {Queiroz}, {Quirrenbach}, {Rahurkar}, {Raichoor}, {Ramstedt},
  {Rau}, {Recio-Blanco}, {Reiss}, {Renaud}, {Revaz}, {Rhode}, {Richard},
  {Richter}, {Rix}, {Robotham}, {Roelfsema}, {Romaniello}, {Rosario},
  {Rothmaier}, {Roukema}, {Ruchti}, {Rupprecht}, {Rybizki}, {Ryde}, {Saar},
  {Sadler}, {Sahl{\'e}n}, {Salvato}, {Sassolas}, {Saunders}, {Saviauk},
  {Sbordone}, {Schmidt}, {Schnurr}, {Scholz}, {Schwope}, {Seifert}, {Shanks},
  {Sheinis}, {Sivov}, {Sk{\'u}lad{\'o}ttir}, {Smartt}, {Smedley}, {Smith},
  {Smith}, {Sorce}, {Spitler}, {Starkenburg}, {Steinmetz}, {Stilz}, {Storm},
  {Sullivan}, {Sutherland}, {Swann}, {Tamone}, {Taylor}, {Teillon}, {Tempel},
  {ter Horst}, {Thi}, {Tolstoy}, {Trager}, {Traven}, {Tremblay}, {Tresse},
  {Valentini}, {van de Weygaert}, {van den Ancker}, {Veljanoski}, {Venkatesan},
  {Wagner}, {Wagner}, {Walcher}, {Waller}, {Walton}, {Wang}, {Winkler},
  {Wisotzki}, {Worley}, {Worseck}, {Xiang}, {Xu}, {Yong}, {Zhao}, {Zheng},
  {Zscheyge}, \& {Zucker}}]{4MOST2019}
{de Jong}, R.~S., {Agertz}, O., {Berbel}, A.~A., {et~al.} 2019, The Messenger,
  175, 3

\bibitem[{{Dey} {et~al.}(2019){Dey}, {Schlegel}, {Lang}, {Blum}, {Burleigh},
  {Fan}, {Findlay}, {Finkbeiner}, {Herrera}, {Juneau}, {Landriau}, {Levi},
  {McGreer}, {Meisner}, {Myers}, {Moustakas}, {Nugent}, {Patej}, {Schlafly},
  {Walker}, {Valdes}, {Weaver}, {Y{\`e}che}, {Zou}, {Zhou}, {Abareshi},
  {Abbott}, {Abolfathi}, {Aguilera}, {Alam}, {Allen}, {Alvarez}, {Annis},
  {Ansarinejad}, {Aubert}, {Beechert}, {Bell}, {BenZvi}, {Beutler}, {Bielby},
  {Bolton}, {Brice{\~n}o}, {Buckley-Geer}, {Butler}, {Calamida}, {Carlberg},
  {Carter}, {Casas}, {Castander}, {Choi}, {Comparat}, {Cukanovaite}, {Delubac},
  {DeVries}, {Dey}, {Dhungana}, {Dickinson}, {Ding}, {Donaldson}, {Duan},
  {Duckworth}, {Eftekharzadeh}, {Eisenstein}, {Etourneau}, {Fagrelius},
  {Farihi}, {Fitzpatrick}, {Font-Ribera}, {Fulmer}, {G{\"a}nsicke},
  {Gaztanaga}, {George}, {Gerdes}, {Gontcho}, {Gorgoni}, {Green}, {Guy},
  {Harmer}, {Hernandez}, {Honscheid}, {Huang}, {James}, {Jannuzi}, {Jiang},
  {Joyce}, {Karcher}, {Karkar}, {Kehoe}, {Kneib}, {Kueter-Young}, {Lan},
  {Lauer}, {Le Guillou}, {Le Van Suu}, {Lee}, {Lesser}, {Perreault Levasseur},
  {Li}, {Mann}, {Marshall}, {Mart{\'\i}nez-V{\'a}zquez}, {Martini}, {du Mas des
  Bourboux}, {McManus}, {Meier}, {M{\'e}nard}, {Metcalfe},
  {Mu{\~n}oz-Guti{\'e}rrez}, {Najita}, {Napier}, {Narayan}, {Newman}, {Nie},
  {Nord}, {Norman}, {Olsen}, {Paat}, {Palanque-Delabrouille}, {Peng},
  {Poppett}, {Poremba}, {Prakash}, {Rabinowitz}, {Raichoor}, {Rezaie},
  {Robertson}, {Roe}, {Ross}, {Ross}, {Rudnick}, {Safonova}, {Saha},
  {S{\'a}nchez}, {Savary}, {Schweiker}, {Scott}, {Seo}, {Shan}, {Silva},
  {Slepian}, {Soto}, {Sprayberry}, {Staten}, {Stillman}, {Stupak}, {Summers},
  {Sien Tie}, {Tirado}, {Vargas-Maga{\~n}a}, {Vivas}, {Wechsler}, {Williams},
  {Yang}, {Yang}, {Yapici}, {Zaritsky}, {Zenteno}, {Zhang}, {Zhang}, {Zhou}, \&
  {Zhou}}]{DESI}
{Dey}, A., {Schlegel}, D.~J., {Lang}, D., {et~al.} 2019, \aj, 157, 168

\bibitem[{{Eisenstein} {et~al.}(2011){Eisenstein}, {Weinberg}, {Agol},
  {Aihara}, {Allende Prieto}, {Anderson}, {Arns}, {Aubourg}, {Bailey},
  {Balbinot}, \& et~al.}]{SDSS_III}
{Eisenstein}, D.~J., {Weinberg}, D.~H., {Agol}, E., {et~al.} 2011, \aj, 142, 72

\bibitem[{{Ganeshaiah Veena} {et~al.}(2019){Ganeshaiah Veena}, {Cautun},
  {Tempel}, {van de Weygaert}, \& {Frenk}}]{Ganeshaiah2019}
{Ganeshaiah Veena}, P., {Cautun}, M., {Tempel}, E., {van de Weygaert}, R., \&
  {Frenk}, C.~S. 2019, \mnras, 487, 1607

\bibitem[{{Klypin} {et~al.}(2016){Klypin}, {Yepes}, {Gottl{\"o}ber}, {Prada},
  \& {He{\ss}}}]{Klypin16}
{Klypin}, A., {Yepes}, G., {Gottl{\"o}ber}, S., {Prada}, F., \& {He{\ss}}, S.
  2016, \mnras, 457, 4340

\bibitem[{{Knebe} {et~al.}(2004){Knebe}, {Gill}, {Gibson}, {Lewis}, {Ibata}, \&
  {Dopita}}]{Knebe2004}
{Knebe}, A., {Gill}, S. P.~D., {Gibson}, B.~K., {et~al.} 2004, \apj, 603, 7

\bibitem[{{Knebe} {et~al.}(2018){Knebe}, {Stoppacher}, {Prada}, {Behrens},
  {Benson}, {Cora}, {Croton}, {Padilla}, {Ruiz}, {Sinha}, {Stevens},
  {Vega-Mart{\'\i}nez}, {Behroozi}, {Gonzalez-Perez}, {Gottl{\"o}ber},
  {Klypin}, {Yepes}, {Enke}, {Libeskind}, {Riebe}, \& {Steinmetz}}]{Knebe18}
{Knebe}, A., {Stoppacher}, D., {Prada}, F., {et~al.} 2018, \mnras, 474, 5206

\bibitem[{{Kraljic} {et~al.}(2020){Kraljic}, {Dav{\'e}}, \&
  {Pichon}}]{Kraljic2020}
{Kraljic}, K., {Dav{\'e}}, R., \& {Pichon}, C. 2020, \mnras, 493, 362

\bibitem[{{Kruuse} {et~al.}(2019){Kruuse}, {Tempel}, {Kipper}, \&
  {Stoica}}]{Kruuse2019}
{Kruuse}, M., {Tempel}, E., {Kipper}, R., \& {Stoica}, R.~S. 2019, \aap, 625,
  A130

\bibitem[{{Kuutma} {et~al.}(2017){Kuutma}, {Tamm}, \& {Tempel}}]{Kuutma2017}
{Kuutma}, T., {Tamm}, A., \& {Tempel}, E. 2017, \aap, 600, L6

\bibitem[{{Laur} {et~al.}(2022){Laur}, {Tempel}, {Tamm}, {Kipper},
  {Liivam{\"a}gi}, {Hern{\'a}n-Caballero}, {Muru}, {Chaves-Montero},
  {D{\'\i}az-Garc{\'\i}a}, {Turner}, {Tuvikene}, {Queiroz}, {Bom},
  {Fern{\'a}ndez-Ontiveros}, {Gonz{\'a}lez Delgado}, {Civera}, {Abramo},
  {Alcaniz}, {Ben{\'\i}tez}, {Bonoli}, {Carneiro}, {Cenarro},
  {Crist{\'o}bal-Hornillos}, {Dupke}, {Ederoclite}, {L{\'o}pez-Sanjuan},
  {Mar{\'\i}n-Franch}, {de Oliveira}, {Moles}, {Sodr{\'e}}, {Taylor}, {Varela},
  \& {Rami{\'o}}}]{Laur2022}
{Laur}, J., {Tempel}, E., {Tamm}, A., {et~al.} 2022, \aap, 668, A8

\bibitem[{{Lee} \& {Pen}(2000)}]{LeePen2000}
{Lee}, J. \& {Pen}, U.-L. 2000, \apjl, 532, L5

\bibitem[{{Libeskind} {et~al.}(2018){Libeskind}, {van de Weygaert}, {Cautun},
  {Falck}, {Tempel}, {Abel}, {Alpaslan}, {Arag{\'o}n-Calvo}, {Forero-Romero},
  {Gonzalez}, {Gottl{\"o}ber}, {Hahn}, {Hellwing}, {Hoffman}, {Jones},
  {Kitaura}, {Knebe}, {Manti}, {Neyrinck}, {Nuza}, {Padilla}, {Platen},
  {Ramachandra}, {Robotham}, {Saar}, {Shand arin}, {Steinmetz}, {Stoica},
  {Sousbie}, \& {Yepes}}]{Libeskind18}
{Libeskind}, N.~I., {van de Weygaert}, R., {Cautun}, M., {et~al.} 2018, \mnras,
  473, 1195

\bibitem[{{Muru} \& {Tempel}(2021)}]{Muru21}
{Muru}, M.~M. \& {Tempel}, E. 2021, \aap, 649, A108

\bibitem[{{Nevalainen} {et~al.}(2015){Nevalainen}, {Tempel}, {Liivam{\"a}gi},
  {Branchini}, {Roncarelli}, {Giocoli}, {Hein{\"a}m{\"a}ki}, {Saar}, {Tamm},
  {Finoguenov}, {Nurmi}, \& {Bonamente}}]{Nevalainen15}
{Nevalainen}, J., {Tempel}, E., {Liivam{\"a}gi}, L.~J., {et~al.} 2015, \aap,
  583, A142

\bibitem[{{Planck Collaboration} {et~al.}(2016){Planck Collaboration}, {Adam,
  R.}, {Ade, P. A. R.}, {Aghanim, N.}, {Akrami, Y.}, {Alves, M. I. R.},
  {Arg\"ueso, F.}, {Arnaud, M.}, {Arroja, F.}, {Ashdown, M.}, {Aumont, J.},
  {Baccigalupi, C.}, {Ballardini, M.}, {Banday, A. J.}, {Barreiro, R. B.},
  {Bartlett, J. G.}, {Bartolo, N.}, {Basak, S.}, {Battaglia, P.}, {Battaner,
  E.}, {Battye, R.}, {Benabed, K.}, {Beno\^{\i}t, A.}, {Benoit-L\'evy, A.},
  {Bernard, J.-P.}, {Bersanelli, M.}, {Bertincourt, B.}, {Bielewicz, P.},
  {Bikmaev, I.}, {Bock, J. J.}, {B\"ohringer, H.}, {Bonaldi, A.}, {Bonavera,
  L.}, {Bond, J. R.}, {Borrill, J.}, {Bouchet, F. R.}, {Boulanger, F.},
  {Bucher, M.}, {Burenin, R.}, {Burigana, C.}, {Butler, R. C.}, {Calabrese,
  E.}, {Cardoso, J.-F.}, {Carvalho, P.}, {Casaponsa, B.}, {Castex, G.},
  {Catalano, A.}, {Challinor, A.}, {Chamballu, A.}, {Chary, R.-R.}, {Chiang, H.
  C.}, {Chluba, J.}, {Chon, G.}, {Christensen, P. R.}, {Church, S.}, {Clemens,
  M.}, {Clements, D. L.}, {Colombi, S.}, {Colombo, L. P. L.}, {Combet, C.},
  {Comis, B.}, {Contreras, D.}, {Couchot, F.}, {Coulais, A.}, {Crill, B. P.},
  {Cruz, M.}, {Curto, A.}, {Cuttaia, F.}, {Danese, L.}, {Davies, R. D.},
  {Davis, R. J.}, {de Bernardis, P.}, {de Rosa, A.}, {de Zotti, G.},
  {Delabrouille, J.}, {Delouis, J.-M.}, {D\'esert, F.-X.}, {Di Valentino, E.},
  {Dickinson, C.}, {Diego, J. M.}, {Dolag, K.}, {Dole, H.}, {Donzelli, S.},
  {Dor\'e, O.}, {Douspis, M.}, {Ducout, A.}, {Dunkley, J.}, {Dupac, X.},
  {Efstathiou, G.}, {Eisenhardt, P. R. M.}, {Elsner, F.}, {En\ss{}lin, T. A.},
  {Eriksen, H. K.}, {Falgarone, E.}, {Fantaye, Y.}, {Farhang, M.}, {Feeney,
  S.}, {Fergusson, J.}, {Fernandez-Cobos, R.}, {Feroz, F.}, {Finelli, F.},
  {Florido, E.}, {Forni, O.}, {Frailis, M.}, {Fraisse, A. A.}, {Franceschet,
  C.}, {Franceschi, E.}, {Frejsel, A.}, {Frolov, A.}, {Galeotta, S.}, {Galli,
  S.}, {Ganga, K.}, {Gauthier, C.}, {G\'enova-Santos, R. T.}, {Gerbino, M.},
  {Ghosh, T.}, {Giard, M.}, {Giraud-H\'eraud, Y.}, {Giusarma, E.}, {Gjerl\o{}w,
  E.}, {Gonz\'alez-Nuevo, J.}, {G\'orski, K. M.}, {Grainge, K. J. B.},
  {Gratton, S.}, {Gregorio, A.}, {Gruppuso, A.}, {Gudmundsson, J. E.}, {Hamann,
  J.}, {Handley, W.}, {Hansen, F. K.}, {Hanson, D.}, {Harrison, D. L.},
  {Heavens, A.}, {Helou, G.}, {Henrot-Versill\'e, S.}, {Hern\'andez-Monteagudo,
  C.}, {Herranz, D.}, {Hildebrandt, S. R.}, {Hivon, E.}, {Hobson, M.}, {Holmes,
  W. A.}, {Hornstrup, A.}, {Hovest, W.}, {Huang, Z.}, {Huffenberger, K. M.},
  {Hurier, G.}, {Ili\'{}c, S.}, {Jaffe, A. H.}, {Jaffe, T. R.}, {Jin, T.},
  {Jones, W. C.}, {Juvela, M.}, {Karakci, A.}, {Keih\"anen, E.}, {Keskitalo,
  R.}, {Khamitov, I.}, {Kiiveri, K.}, {Kim, J.}, {Kisner, T. S.}, {Kneissl,
  R.}, {Knoche, J.}, {Knox, L.}, {Krachmalnicoff, N.}, {Kunz, M.},
  {Kurki-Suonio, H.}, {Lacasa, F.}, {Lagache, G.}, {L\"ahteenm\"aki, A.},
  {Lamarre, J.-M.}, {Langer, M.}, {Lasenby, A.}, {Lattanzi, M.}, {Lawrence, C.
  R.}, {Le Jeune, M.}, {Leahy, J. P.}, {Lellouch, E.}, {Leonardi, R.},
  {Le\'on-Tavares, J.}, {Lesgourgues, J.}, {Levrier, F.}, {Lewis, A.},
  {Liguori, M.}, {Lilje, P. B.}, {Lilley, M.}, {Linden-V\o{}rnle, M.},
  {Lindholm, V.}, {Liu, H.}, {L\'opez-Caniego, M.}, {Lubin, P. M.}, {Ma,
  Y.-Z.}, {Mac\'{\i}as-P\'erez, J. F.}, {Maggio, G.}, {Maino, D.}, {Mak, D. S.
  Y.}, {Mandolesi, N.}, {Mangilli, A.}, {Marchini, A.}, {Marcos-Caballero, A.},
  {Marinucci, D.}, {Maris, M.}, {Marshall, D. J.}, {Martin, P. G.},
  {Martinelli, M.}, {Mart\'{\i}nez-Gonz\'alez, E.}, {Masi, S.}, {Matarrese,
  S.}, {Mazzotta, P.}, {McEwen, J. D.}, {McGehee, P.}, {Mei, S.}, {Meinhold, P.
  R.}, {Melchiorri, A.}, {Melin, J.-B.}, {Mendes, L.}, {Mennella, A.},
  {Migliaccio, M.}, {Mikkelsen, K.}, {Millea, M.}, {Mitra, S.},
  {Miville-Desch\^enes, M.-A.}, {Molinari, D.}, {Moneti, A.}, {Montier, L.},
  {Moreno, R.}, {Morgante, G.}, {Mortlock, D.}, {Moss, A.}, {Mottet, S.},
  {M\"unchmeyer, M.}, {Munshi, D.}, {Murphy, J. A.}, {Narimani, A.}, {Naselsky,
  P.}, {Nastasi, A.}, {Nati, F.}, {Natoli, P.}, {Negrello, M.}, {Netterfield,
  C. B.}, {N\o{}rgaard-Nielsen, H. U.}, {Noviello, F.}, {Novikov, D.},
  {Novikov, I.}, {Olamaie, M.}, {Oppermann, N.}, {Orlando, E.}, {Oxborrow, C.
  A.}, {Paci, F.}, {Pagano, L.}, {Pajot, F.}, {Paladini, R.}, {Pandolfi, S.},
  {Paoletti, D.}, {Partridge, B.}, {Pasian, F.}, {Patanchon, G.}, {Pearson, T.
  J.}, {Peel, M.}, {Peiris, H. V.}, {Pelkonen, V.-M.}, {Perdereau, O.},
  {Perotto, L.}, {Perrott, Y. C.}, {Perrotta, F.}, {Pettorino, V.},
  {Piacentini, F.}, {Piat, M.}, {Pierpaoli, E.}, {Pietrobon, D.},
  {Plaszczynski, S.}, {Pogosyan, D.}, {Pointecouteau, E.}, {Polenta, G.},
  {Popa, L.}, {Pratt, G. W.}, {Pr\'ezeau, G.}, {Prunet, S.}, {Puget, J.-L.},
  {Rachen, J. P.}, {Racine, B.}, {Reach, W. T.}, {Rebolo, R.}, {Reinecke, M.},
  {Remazeilles, M.}, {Renault, C.}, {Renzi, A.}, {Ristorcelli, I.}, {Rocha,
  G.}, {Roman, M.}, {Romelli, E.}, {Rosset, C.}, {Rossetti, M.}, {Rotti, A.},
  {Roudier, G.}, {Rouill\'e d\'{}Orfeuil, B.}, {Rowan-Robinson, M.},
  {Rubi\~no-Mart\'{\i}n, J. A.}, {Ruiz-Granados, B.}, {Rumsey, C.}, {Rusholme,
  B.}, {Said, N.}, {Salvatelli, V.}, {Salvati, L.}, {Sandri, M.}, {Sanghera, H.
  S.}, {Santos, D.}, {Saunders, R. D. E.}, {Sauv\'e, A.}, {Savelainen, M.},
  {Savini, G.}, {Schaefer, B. M.}, {Schammel, M. P.}, {Scott, D.}, {Seiffert,
  M. D.}, {Serra, P.}, {Shellard, E. P. S.}, {Shimwell, T. W.}, {Shiraishi,
  M.}, {Smith, K.}, {Souradeep, T.}, {Spencer, L. D.}, {Spinelli, M.},
  {Stanford, S. A.}, {Stern, D.}, {Stolyarov, V.}, {Stompor, R.}, {Strong, A.
  W.}, {Sudiwala, R.}, {Sunyaev, R.}, {Sutter, P.}, {Sutton, D.}, {Suur-Uski,
  A.-S.}, {Sygnet, J.-F.}, {Tauber, J. A.}, {Tavagnacco, D.}, {Terenzi, L.},
  {Texier, D.}, {Toffolatti, L.}, {Tomasi, M.}, {Tornikoski, M.}, {Tramonte,
  D.}, {Tristram, M.}, {Troja, A.}, {Trombetti, T.}, {Tucci, M.}, {Tuovinen,
  J.}, {T\"urler, M.}, {Umana, G.}, {Valenziano, L.}, {Valiviita, J.}, {Van
  Tent, F.}, {Vassallo, T.}, {Vibert, L.}, {Vidal, M.}, {Viel, M.}, {Vielva,
  P.}, {Villa, F.}, {Wade, L. A.}, {Walter, B.}, {Wandelt, B. D.}, {Watson,
  R.}, {Wehus, I. K.}, {Welikala, N.}, {Weller, J.}, {White, M.}, {White, S. D.
  M.}, {Wilkinson, A.}, {Yvon, D.}, {Zacchei, A.}, {Zibin, J. P.}, \& {Zonca,
  A.}}]{Planck15}
{Planck Collaboration}, {Adam, R.}, {Ade, P. A. R.}, {et~al.} 2016, A\&A, 594,
  A1

\bibitem[{{Ruiz-Macias} {et~al.}(2021){Ruiz-Macias}, {Zarrouk}, {Cole},
  {Baugh}, {Norberg}, {Lucey}, {Dey}, {Eisenstein}, {Doel}, {Gazta{\~n}aga},
  {Hahn}, {Kehoe}, {Kitanidis}, {Landriau}, {Lang}, {Moustakas}, {Myers},
  {Prada}, {Schubnell}, {Weinberg}, \& {Wilson}}]{DESI_BGS}
{Ruiz-Macias}, O., {Zarrouk}, P., {Cole}, S., {et~al.} 2021, \mnras, 502, 4328

\bibitem[{{Sousbie}(2011)}]{Sousbie2011}
{Sousbie}, T. 2011, \mnras, 414, 350

\bibitem[{{Taylor}(2005)}]{TOPCAT}
{Taylor}, M.~B. 2005, in Astronomical Society of the Pacific Conference Series,
  Vol. 347, Astronomical Data Analysis Software and Systems XIV, ed.
  P.~{Shopbell}, M.~{Britton}, \& R.~{Ebert}, 29

\bibitem[{{Tempel} {et~al.}(2015){Tempel}, {Guo}, {Kipper}, \&
  {Libeskind}}]{Tempel2015}
{Tempel}, E., {Guo}, Q., {Kipper}, R., \& {Libeskind}, N.~I. 2015, \mnras, 450,
  2727

\bibitem[{{Tempel} \& {Libeskind}(2013)}]{TempelLibeskind2013}
{Tempel}, E. \& {Libeskind}, N.~I. 2013, \apjl, 775, L42

\bibitem[{{Tempel} {et~al.}(2016){Tempel}, {Stoica}, {Kipper}, \&
  {Saar}}]{Tempel16}
{Tempel}, E., {Stoica}, R.~S., {Kipper}, R., \& {Saar}, E. 2016, Astronomy and
  Computing, 16, 17

\bibitem[{{Tempel} {et~al.}(2014){Tempel}, {Stoica}, {Mart{\'\i}nez},
  {Liivam{\"a}gi}, {Castellan}, \& {Saar}}]{Tempel14}
{Tempel}, E., {Stoica}, R.~S., {Mart{\'\i}nez}, V.~J., {et~al.} 2014, \mnras,
  438, 3465

\bibitem[{{Tuominen} {et~al.}(2021){Tuominen}, {Nevalainen}, {Tempel},
  {Kuutma}, {Wijers}, {Schaye}, {Hein{\"a}m{\"a}ki}, {Bonamente}, \&
  {Ganeshaiah Veena}}]{Tuominen2021}
{Tuominen}, T., {Nevalainen}, J., {Tempel}, E., {et~al.} 2021, \aap, 646, A156

\bibitem[{{University of Tartu}(2018)}]{ut_hpc}
{University of Tartu}. 2018, UT Rocket

\bibitem[{van~der Plas {et~al.}(2022)van~der Plas, Dral, Berg,
  Γεωργακόπουλος, Huijzer, Bochenski, Mengali, Lungwitz, Burns,
  Priyashan, Ling, Zhang, Schneider, Weaver, Rogerluo, Kadowaki, Wu, Gerritsen,
  Novosel, Supanat, Moon, pupuis, Abbott, Bauer, Bouffard, Terasaki, Polasa, \&
  TheCedarPrince}]{julia_pluto}
van~der Plas, F., Dral, M., Berg, P., {et~al.} 2022, fonsp/Pluto.jl: v0.19.11

\bibitem[{{Wang} {et~al.}(2020){Wang}, {Libeskind}, {Tempel}, {Pawlowski},
  {Kang}, \& {Guo}}]{Wang2020}
{Wang}, P., {Libeskind}, N.~I., {Tempel}, E., {et~al.} 2020, \apj, 900, 129

\bibitem[{White {et~al.}(2020)White, Kamiński, powerdistribution,
  Bouchet-Valat, Garborg, Quinn, Kornblith, cjprybol, Stukalov, Bates, Short,
  DuBois, Harris, Squire, Arslan, pdeffebach, Anthoff, Kleinschmidt, Noack,
  Shah, Mellnik, Arakaki, Mohapatra, Peter, Karpinski, Lin, timema,
  ExpandingMan, Oswald, \& White}]{julia_dataframes}
White, J.~M., Kamiński, B., powerdistribution, {et~al.} 2020,
  JuliaData/DataFrames.jl: v0.22.1

\bibitem[{{Zentner} {et~al.}(2005){Zentner}, {Kravtsov}, {Gnedin}, \&
  {Klypin}}]{Zenter2005}
{Zentner}, A.~R., {Kravtsov}, A.~V., {Gnedin}, O.~Y., \& {Klypin}, A.~A. 2005,
  \apj, 629, 219

\end{thebibliography}

\end{document}